\documentclass[epsfig]{kluwer}    

\newdisplay{guess}{Conjecture}

\begin{document}                                                                                   
\begin{article}
\begin{opening}         
\title{Inner Disk Oscillations and QPOs in Relativistic Jet Sources}
\author{A. R. \surname{Rao}}  
\runningauthor{A. R. Rao}
\runningtitle{Inner disk oscillations and QPOs}
\institute{Tata Institute of Fundamental Research, Mumbai, INDIA }
\date{October 15, 2000}

\begin{abstract}
Recent results on the inner disk oscillations found in 
GRS 1915+105 are reviewed. QPOs during the low state are used 
as a marker for such oscillations and the physical picture emerging 
from a combined X-ray spectral and timing analysis is
examined. The relationship between inner disk oscillations and
synchrotron radio emission is critically evaluated.
\end{abstract}

\end{opening}           

\section{Introduction }  

One of the remarkable features of the X-ray variability of
the Galactic relativistic jet source GRS~1915+105 is the 
rapid intensity variations associated with repeatable and
distinct spectral characteristics.
Belloni et al. (1997) found that
the source shows distinct and different inner disk radius 
during the ``quiescence'' and outburst which can be interpreted as
the vanishing of the  bright inner disk during the quiescence. 
Detection of  rapid radio oscillations near the onset of 
superluminally moving ejecta prompted Fender et al. (1999) to
suggest that such X-ray inner disk oscillations could be the 
underlying reason for the ejection of the synchrotron cloud responsible
for the radio emission.

Ever since its discovery (Castro-Tirado et al. 1992) GRS~1915+105 has
been exhibiting a rich variety of X-ray variability characteristics.
Greiner et al. (1996) presented the RXTE observations of the source
which included repeated patterns
of brightness sputters, large amplitude oscillations, fast oscillations
and several incidences of prolonged lulls.
Belloni et al. (1997) classified some of the oscillatory
X-ray variations as ``outbursts''  and from a time resolved
spectral analysis found evidence for the disappearance of the
inner disk during these oscillations.
They also discovered a strong correlation between the quiescent
phase and burst duration. 
Taam, Chen, and Swank (1997)
detected a wide range of transient activity including regular bursts with
a recurrence time of about one minute and irregular bursts. 
Paul et al. (1998) detected several
types of such bursts using the IXAE data and found evidence for matter
disappearing into the event horizon of the black hole. Yadav et al. (1999)
made a systematic analysis of these bursts and classified them based
on the recurrence time.

 Belloni et al. (2000a) made a detailed evaluation of the X-ray 
variability characteristics of GRS~1915+105 and classified the
variations in 12 different classes. To understand the basic nature of these
classes, 
in Figure 1 we give the representative light curves and hardness
ratios obtained from the {\it RXTE} archives. The total count rate, R,
is in the units of 10$^4$ counts s$^{-1}$ and it is vertically
shifted by 5 units. The hardness ratio H1 (5 -- 13 keV to 2 -- 5 keV
ratio) is scaled by 5 and shifted by 2 units and the hardness ratio H2
($>$ 13 keV to 2 -- 13 keV ratio) is scaled up by 10 and shifted by
4 units. The hardness ratio H1 is a measure of the disk blackbody
temperature and H2 is a measure of the intensity of the hard power-law,
presumably arising from a thermal-Compton cloud.

\begin{figure} 
\vspace{30pc}
\includegraphics{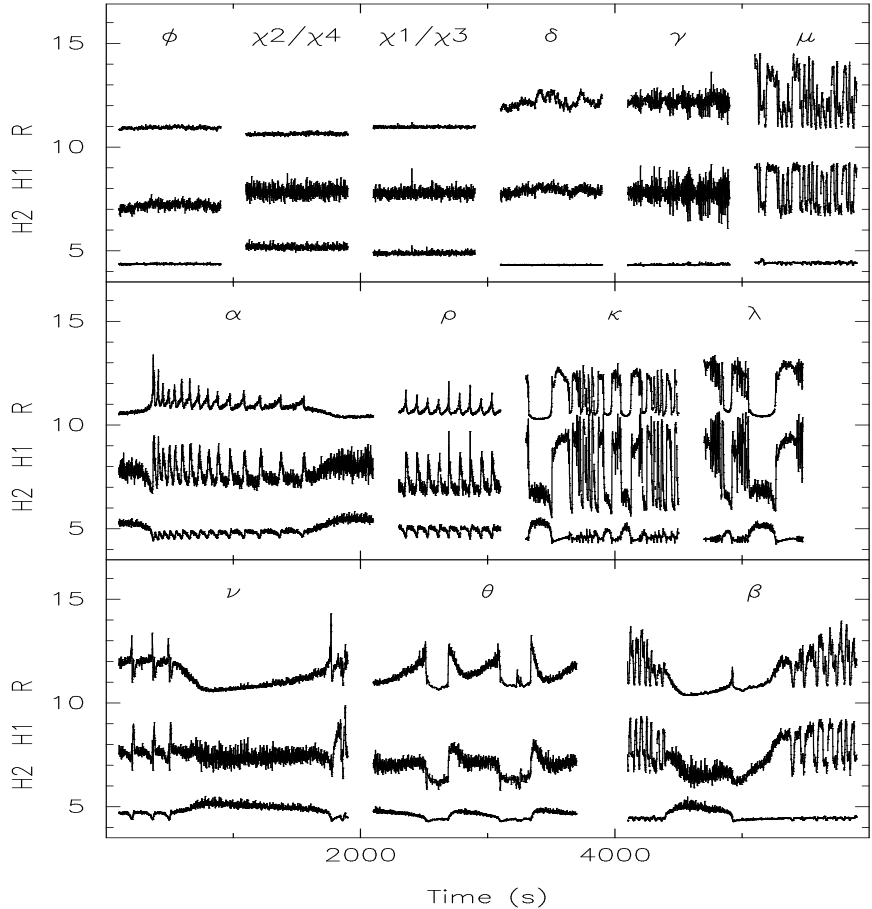}
\caption[]{X-ray count rates (R) and the hardness ratios for the various classes
of variability seen in GRS~1915+105.}
\label{claSS}
\end{figure}

 The  X-ray variability classes which show steady behavior during one
orbit of RXTE are shown in the top panel of the figure. Class $\phi$
is characterized by low count rate, low H1 and very low H2. It is 
also devoid of any strong variability. Class $\chi$ is characterized
by strong 0.5 -- 10 Hz QPO and it resembles the intermediate state
of other black hole sources (Trudolyubov  et al. 1999b). Classes 
$\delta$, $\gamma$ and $\mu$ are high-soft states of the source
and occasionally the 67 Hz QPOs are seen in these states (Morgan et al.
1997).

 The various X-ray variability classes which exhibit inner disk
oscillations are shown in the middle panel of Figure 1. The distinguishing
feature is the sharp ($<$ 10 s) change in the intensity state. During the
high state of these oscillations the source exhibits all the temporal
and spectral characteristics of the high-soft state of the source and
during the low state it exhibits all characteristics of the low-hard state
of the source (Rao et al. 2000b). The drastic change in  H2,
signifying a change in the Compton cloud, is a common feature of 
this class. The behavior can be regular (class $\rho$) and it can
remain in this class for periods up to several tens of days. In 
 class $\alpha$ the source  
reverts back and forth between class $\chi$ and class $\rho$.
During the quiescent periods of all these classes the 0.5 -- 10 Hz
QPO is always present.

 In contrast to such inner disk oscillations, classes $\nu$, $\theta$,
and $\beta$ show a more gradual change from a high-soft to low-hard 
state and they are characterized
by a sharp soft dip when both H1 and H2 are low and the variability too
drops to a low level. Radio oscillations are associated with these
classes (Eikenberry et al. 1998; Mirabel et al. 1998).

\section{Inner Disk Oscillations and QPOs }

 One of the remarkable features of the inner disk oscillations is the
presence of the ubiquitous 0.5 -- 10 Hz QPO during the quiescent state
of the oscillations. This type of  QPO is found during the low-hard state of the
source (and also in the low-hard state of other jet sources as well), 
and it is used as a marker for the low-hard state of the source
(Muno et al. 1999; Trudolyubov  et al. 1999b). 

Reig et al. (2000) made a detailed study of the QPO phenomena
in the $\chi$1 and $\chi$3 classes
and found a correlation between the QPO frequency and the
observed phase lag in the Fourier cross-spectrum in two energy
channels. Nobili et al. (2000) have interpreted this
correlation as due to arising from the geometry of
the region producing the Compton spectrum, assumed to be arising
from a corona formed by the puffed up inner
regions of the accretion disk. 
Rao et al. (2000a) made a detailed spectral study during
the $\chi$3 state of the source and identified three
spectral components. They found that the QPOs are due to the
Compton component and they invoked an accretion disk shock to explain
the QPO phenomena.

 The large rms amplitude (about 10\%), narrow width and the relative stability
over considerably long durations makes it difficult to explain
the 0.5 -- 10 Hz QPO on the basis of any disk oscillation model.
Chakrabarti \& Manickam (2000) 
interpreted the QPOs  as due to the oscillation of
the
region responsible for the hard radiation, i.e., the post-shock region.
They also detected a correlation  between the average frequency of the
QPO and the duration of the `off' states. A similar correlation with
centroid frequencies were
also found by Trudolyubov, Churazov, \& Gilfanov (1999a).


There have been attempts to explain the rapid variability seen in
GRS 1915+105 using disk instability models. 
Belloni et al. (1997)
have tried to explain the repeated patterns as due to the appearance
and disappearance of the inner accretion disk.
Nayakshin et al. (1999)  have investigated the different accretion models and
viscosity prescriptions and attempted to explain the temporal behavior of
GRS 1915+105. In particular, they have shown that the accretion instability
in a slim disk is not likely to
adequately account for the behavior of GRS 1915+105 (because of the difficulty
in maintaining the high state for long) and the appearance of the inner
accretion disk, as postulated by Belloni et al. (1997), will
take a time scale comparable to the burst time scale. Though Nayakshin
et al. (1999) were able to reproduce many characteristics in the
X-ray variability for GRS~1915+105, they were unable to explain the
rise/fall times or the $<$ 10 s oscillations.

\begin{figure} 
\vspace{17pc}
\includegraphics{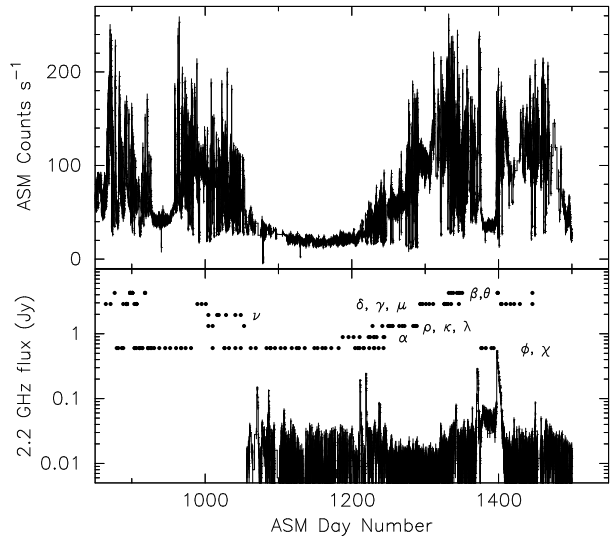}
\caption[]{The ASM count rates for GRS 1915+105 is shown in the top
panel and the 2.2 GHz GBI data is shown in the bottom panel. The time of
occurrence of different intensity classes as established from the
pointed RXTE PCA observations (Belloni et al. 2000a) are also shown in the bottom panel
of the figure.
}
\label{radiO}
\end{figure}

 Yadav et al. (1999)
presented a  comprehensive picture for the origin  of these  bursts. They
suggested that the peculiar bursts
are characteristic of the change of state of the source.
The source can switch back and forth between the low-hard state and
the high-soft state  near  critical accretion rates in a very short time
scale, giving rise to the irregular and quasi-regular  bursts.  
It was pointed out that changes in total accretion
rate cannot  manifest itself in short time scales and hence the
fast spectral state changes in GRS~1915+105 cannot be explained by
changes in total accretion rates.  Chakrabarti \& Manickam (2000)
provided an explanation for this remarkable
observation by invoking mass loss. Unlike the canonical low-hard state
where the hard power-law is due to Comptonisation, in the hard state
observed during the bursts there should be an additional feature due
to winds. Chakrabarti et al. (2000) have observed 
subtle differences in the spectra 
during the bursts compared to the spectra during the long duration low-hard
and high-soft states.
This change is above $\sim$ 10 keV where the hard component
of the low state softens. This is the
effect predicted in Chakrabarti (1998) where the addition of about
10\% mass loss in wind has an effect of pushing the spectral cross-over
point from 15-17 keV (for a 10 M$_\odot$ black hole) to higher energies.

\section{Inner Disk Oscillations and Synchrotron Radio Emission}

There have been attempts to connect the X-ray variability 
characteristics to the onset of jet emission in GRS~1915+105.
Fender et al. (1999) made a detailed analysis of the super-luminal jet ejection
events observed in GRS 1915$+$105 in 1997 October/November and detected
continuous short period (20$-$40 minute) radio oscillations during the
start of the jet emission. They proposed that these are indications of
repeated ejection of the inner accretion disk, quite similar to the events
seen in X-rays by Belloni et al. (1997). A causal connection between disk
and jet was thus attempted.
 Naik et al. (2000) have reported the detection of a series
of X-ray dips during a huge radio flare of strength 0.48 Jy (at 2.25 GHz).
They argue that a large number of such X-ray dips can account for the
radio flare emission.

 To put such disk-jet connections in perspective, we have shown in
Figure  2 (top panel) the observed ASM count rates during 1996/1997.
The concurrently obtained RXTE-PCA data have been classified 
by Belloni et al. (2000a) and these are shown in the bottom panel of the
figure, along with the 2.2 GHz radio data obtained from GBI public domain.
The 1997 slow state transition is particularly interesting (see also
Rao et al. 2000b).

It can be seen from the figure  that the source was in a stable low-hard state
up to 1997 April 25 (ASM day number 1210).
The radio flux during this state is 20 -- 30 mJy. This is
the radio-quiet low-hard state of the source (Muno et al. 1999).
The source started a steady increase in its
X-ray emission with an average increase in the ASM count rate
of 0.65 s$^{-1}$ day$^{-1}$ and reaching a count rate of 76 s$^{-1}$
in the middle of July (day number 1290). 
During the beginning of the transition, the source showed 
several episodes of $\alpha$ states, accompanied by brief 
radio flares. For more than a month the source showed
several episodes of the inner disk oscillations (classes $\rho$,
$\kappa$ and $\lambda$)
 (Yadav et al. 1999; Belloni et al. 1997).
 The source reached a steady high-soft state (classes $\delta$, 
$\gamma$ and $\mu$) when the radio emission is very low
($\le$ 20 mJy). 
The ringing flares started again in
the beginning of 1997 August (Yadav et al. 1999) and towards the end of
this state the peculiar outbursts (class $\beta$) accompanied by infrared flares
were observed on 1997 August 14-15 (Eikenberry et al. 1998).
On September 9 the peculiar
``outburst'' was seen again (Markwardt et al. 1999).
In October 1997 the source reached the radio-loud hard-steady state
when both the X-ray and radio fluxes were high without
the evidence of any oscillations.
Fender et al. (1999) described this class as the ``plateau'' state.

From these we can conclude that the inner disk oscillations
involving rapid transitions (classes $\rho$, $\kappa$ and $\lambda$)
do not lead to any appreciable radio emission. Apart from the 
radio-loud hard state, classes $\theta$ and $\beta$ are associated
with enhanced radio emission.
Class $\beta$ are associated with the synchrotron flares in
radio (Mirabel et al. 1998; Fender \& Pooley 1998) and infrared
(Eikenberry et al. 1998). From simultaneous X-ray and infrared
observations, Eikenberry et al. (1998) made a strong argument that
the onset of radio/infrared flare is associated with the soft dip
rather than the gradual change to the low-hard state.
Radio oscillations on a time scale of
20 $-$ 30 minutes are seen to be accompanied by a series of soft X-ray
dips (Fig. 10, Dhawan et al. 2000), similar to the oscillations observed
by Fender et al. (1999). Since the soft dip events are associated with
the jet emission, Naik et al. (2000) proposed that the huge
radio flares are produced by a series of such soft X-ray dips.

  Naik and Rao (2000) made a systematic study of the radio and 
X-ray emissions from GRS 1915+105 and found an one-to-one association
between radio flares and the $\beta$ and $\theta$ classes.
If the onset of a huge radio flare (signifying the emission
of a superluminal ejecta) is associated with an X-ray emission
characteristic observed so far, it has to be necessarily 
with the $\beta$ or $\theta$ classes. Belloni et al. 
(2000b), however, associate the change of state with the radio
emission rather than the peculiar dips seen in these two classes.
A continuous X-ray monitoring during a radio flare should clarify
this  question.

\acknowledgements

I  thank the members of the RXTE and NSF-NRAO-NASA GBI team for making the
data publicly available. 
This research has made use of data obtained through the High Energy
Astrophysics Science Archive Research Center Online Service, provided by the
NASA/Goddard Space Flight Center.
I also thank S. Naik and  S. V. Vadawale for helpful discussions
and assistance in preparing this manuscript.

\end{article}

\begin{thebibliography}{}

\bibitem[Belloni et al. (1997a)]{bell:97a}
Belloni, T., M. Mendez, A.R. King, M. van der Klis and
J. van Paradijs. An Unstable Central Disk in the Superluminal Black Hole X-ray
Binary GRS 1915+105.  
{\it   ApJ}, 479:L145--L148, 1997


\bibitem[Belloni et al. (2000a)]{bell:00a}
Belloni, T., M. Klein-Wolt, M. Mendez, M. van der Klis and J.  van Paradijs.
A model-independent analysis of the variability of GRS 1915+105. {\it
 A\&A},  355:271-290, 2000a.

\bibitem[Belloni et al. 2000b]{bell:00b}
Belloni, T., S. Migliari, S. and R. P. Fender.
Disk Mass Accretion Rate and Infrared Flares in GRS 1915+105.
{\it A\&A}, 358:L29--L32, 2000b.

\bibitem[Castro-Tirado et al. 1992]{cast:92}
Castro-Tirado, A.J., S. Brandt and N.  Lund. {\it IAU Circ.}, 5590, 1992.

\bibitem[Chakrabarti 1998]{chak:98}
Chakrabarti, S. K. Spectral Softening Due to Winds in Accretion Disks.
{\em Indian Journal of Physics}, 72B(6):565--569, 1998

\bibitem[Chakrabarti et al. 2000]{chak:00}
Chakrabarti, S. K., S. G. Manickam, A. K. Nandi and A. R.  Rao.
Spectral Signature of Wind Formation from Post-shock Region
in GRS 1915+105 Accretion Disk.
{\it A\&A}, submitted,  2000.

\bibitem[Chakrabarti \& Manickam  (1999)]{chak:99a}
Chakrabarti, S. K. and S. G. Manickam. 
Correlation among Quasi-periodic Oscillation Frequencies and 
Quiescent-state Duration in Black Hole Candidate GRS 1915+105.
{\it ApJ}, 531:L41--L44, 2000.


\bibitem[Dhawan et al. 2000]{dhaw:00}
Dhawan, V., I. F.  Mirabel and L. F. Rodriguez. 
AU-Scale Synchrotron Jets and Superluminal Ejecta in GRS 1915+105.
 {\it ApJ}, 543:373--385, 2000.

\bibitem[Eikenberry et al. 1998]{eiken:98}
Eikenberry, S. S., K.  Matthews, E. H. Morgan, R. A. 
Remillard and R. W. Nelson. Evidence for a Disk-Jet Interaction in the 
Microquasar GRS 1915+105. {\it ApJ}, 494:L61--L64,  1998.


\bibitem[Fender et al. 1999]{fend:99}
Fender, R. P., S. T. Garrington, D. J.  McKay,  T. W. B. Muxlow,
R. E. Spencer, A. M. Sterling and E. B. Waltman. 
MERLIN observations of relativistic ejections from GRS 1915+105.
{\it
MNRAS}, 304:865--876, 1999.

\bibitem[Fender \& Pooley, 2000]{fp:98}
Fender, R. P. and G. G. Pooley.
Infrared synchrotron oscillations in GRS 1915+105. {\it 
MNRAS}, 300:573--576, 1998.

\bibitem[Greiner et al, 2000]{gra:96}
Greiner, J., E. H.  Morgan and R. A.  Remillard. 
Rossi X-Ray Timing Explorer Observations of GRS 1915+105. 
{\it ApJ}, 473:L107--L110, 1996.

\bibitem[Markwardt et al. 1999]{mark:99}
Markwardt, C.B., J.H. Swank and R.E.  Taam.
Variable-frequency Quasi-periodic Oscillations from the
Galactic Microquasar GRS 1915+105. {\it ApJ}, 513:L37--L40, 1999.

\bibitem[Mirabel \& Rodriguez 1994]{mira:94}
Mirabel, I. F., V. Dhawan, S. Chaty, L. F. Rodriguez, J. Marti,
C. R. Robinson, J. Swank and  T. Geballe. Accretion instabilities and jet formation in GRS 1915+105. {\it  A\&A}, 330:L9--L12, 1998.

\bibitem[Morgan et al. 1997]{morg:97}
Morgan, E. H., R. A.  Remillard and J.  Greiner. 
RXTE Observations of QPOs in the Black Hole Candidate GRS 1915+105.
{\it  ApJ}, 482:993--1010, 1997.

\bibitem[Muno et al. 1999]{muno:99}
Muno, M. P., E.H. Morgan and R. A. Remillard.
Quasi-periodic Oscillations and Spectral States in GRS 1915+105.
 {\it ApJ}, 527:321--340, 1999.

\bibitem[Naik \& Rao, 2000]{nr:00}
Naik, S. and A. R.  Rao. 
 Disk$-$jet connection in GRS 1915$+$105: X-ray soft dips as cause
of radio flares. {\t A\&A}, in press,   astro-ph/0008433, 2000.

\bibitem[Naik et al, 2000]{naik:00}
Naik, S., P. C. Agrawal, A. R.  Rao, B.  Paul, S. Seetha and
K. Kasturirangan. 
Detection of a Series of X-ray Dips Coincident with a Radio
Flare in GRS 1915+105.
{\it ApJ},  in press, 
astro-ph/0008414, 2000.

\bibitem[Nayakshin et al, 2000]{nay:00}
Nayakshin, S., S. Rappaport  and F. Melia.
Time-dependent Disk Models for the Microquasar GRS 1915+105.
{\it  ApJ}, 535:798--814, 2000.

\bibitem[Nobili et al, 2000]{nobili:00}
Nobili, L., R. Turolla, L. Zampieri and T.  Belloni. 
A Comptonization Model for Phase-Lag Variability in GRS 1915+105.
{\it  ApJ}, 538:L137--L140, 2000.

\bibitem[Paul et al, 1998]{nobili:98}
Paul, B., P. C. Agrawal, A. R. Rao, M. N. Vahia, J. S. Yadav, S.
Seetha and K. Kasturirangan. 
Quasi-regular X-Ray Bursts from GRS 1915+105 Observed with the
IXAE: Possible Evidence for Matter Disappearing into the Event Horizon
of the Black Hole.  {\it ApJ},  492:L63--L66, 1998.


\bibitem[Rao et al, 2000]{rao:00}
Rao, A. R., S. Naik, S. V.  Vadawale and S. K. Chakrabarti. 
X-ray Spectral Components in the Hard State of GRS 1915+105: Origin of the
0.5 - 10 Hz QPO.
{\it  A\&A}, 
360:L25--L29, 2000a.

\bibitem[Rao, Yadav \& Paul, 2000]{ryp:00}
Rao, A. R., J. S. Yadav and  B. Paul. 
Rapid State Transitions in the Galactic Black Hole Candidate Source GRS 
1915+105. {\it ApJ} in press,  astro-ph/0007250, 2000b.

\bibitem[Reig et al, 2000]{reig:00}
Reig, P., T. Belloni, M. van der Klis, M. Mendez, N. D. Kylafis and E. C. 
Ford. 
Phase Lag Variability Associated with the 0.5--10 Hz Quasi-Periodic Oscillations in GRS 1915+105. {\it 
ApJ}, 541:883--888, 2000

\bibitem[Taam et al, 1997]{taam:97}
Taam, R. E., X. Chen and J. H. Swank. Rapid Bursts from GRS 1915+105 with
RXTE {\it ApJ}, 485:L83--L86, 1997.

\bibitem[Trudolyubov  et al. 1999a]{trud:99a}
Trudolyubov, S., E. Churazov and M.  Gilfanov. 
The 1--12 Hz QPOs and Dips in GRS 1915+105: Tracers of Keplerian 
and Viscous Time Scales ?
{\it A\&A}, 351:L15-L18, 1999a.

\bibitem[Trudolyubov  et al. 1999b]{trud:99b}
Trudolyubov, S., E. Churazov and M.  Gilfanov. 
The X-ray source GRS 1915 + 105: The low-luminosity state and
transitions between the states during 1996-1997 (RXTE observations).
{\it  Ast. L.}, 25:718--738, 1999b.


\bibitem[Yadav et al. 1999]{yadav:99}
Yadav, J. S., A. R. Rao, P. C.  Agrawal, B.  Paul, S. Seetha and K. 
Kasturirangan. 
Different Types of X-ray Bursts from GRS 1915+105 and Their
Origin. {\it ApJ}, 517:935--950, 1999.

\end{thebibliography}
\end{document}